\documentclass[10pt,conference]{IEEEtran}
\IEEEoverridecommandlockouts
\usepackage{cite}
\usepackage{amsmath,amssymb,amsfonts}
\usepackage{algorithmic}
\usepackage{graphicx}
\usepackage{textcomp}
\usepackage{xcolor}
\usepackage{multirow}
\usepackage[backref]{hyperref}
\usepackage{setspace}
\def\BibTeX{{\rm B\kern-.05em{\sc i\kern-.025em b}\kern-.08em
    T\kern-.1667em\lower.7ex\hbox{E}\kern-.125emX}}

\begin{document}
\title{Let's Chat to Find the APIs: Connecting Human, LLM and Knowledge Graph through AI Chain}

\author{\IEEEauthorblockN{1\textsuperscript{st} Qing~Huang}
\IEEEauthorblockA{\textit{Jiangxi Normal University} \\
Nanchang, China \\
qh@whu.edu.cn}
\and
\IEEEauthorblockN{2\textsuperscript{nd} Zhenyu~Wan}
\IEEEauthorblockA{\textit{Jiangxi Normal University} \\
Nanchang, China \\
wanzy@jxnu.edu.cn}
\and
\IEEEauthorblockN{3\textsuperscript{rd} Zhenchang~Xing}
\IEEEauthorblockA{\textit{CSIRO's Data61 \& Australian National University} \\
Canberra, Australian \\
zhenchang.xing@data61.csiro.au}
\and
\IEEEauthorblockN{4\textsuperscript{th} Changjing~Wang}
\IEEEauthorblockA{\textit{Jiangxi Normal University} \\
Nanchang, China \\
wcj@jxnu.edu.cn}
\and
\IEEEauthorblockN{5\textsuperscript{th} Jieshan~Chen}
\IEEEauthorblockA{\textit{CSIRO's Data61} \\
Canberra, Australian \\
jieshan.chen@data61.csiro.au}
\and
\IEEEauthorblockN{6\textsuperscript{th} Xiwei~Xu}
\IEEEauthorblockA{\textit{CSIRO's Data61} \\
Canberra, Australian \\
xiwei.xu@data61.csiro.au}
\and
\IEEEauthorblockN{7\textsuperscript{th} Qinghua~Lu}
\IEEEauthorblockA{\textit{CSIRO's Data61} \\
Canberra, Australian \\
qinghua.lu@data61.csiro.au}
}

\maketitle

\begin{abstract}

API recommendation methods have evolved from literal and semantic keyword matching to query expansion and query clarification. 
The latest query clarification method is knowledge graph (KG)-based, but limitations include out-of-vocabulary (OOV) failures and rigid question templates. 
To address these limitations, we propose a novel knowledge-guided query clarification approach for API recommendation that leverages a large language model (LLM) guided by KG. 
We utilize the LLM as a neural knowledge base to overcome OOV failures, generating fluent and appropriate clarification questions and options. 
We also leverage the structured API knowledge and entity relationships stored in the KG to filter out noise, and transfer the optimal clarification path from KG to the LLM, increasing the efficiency of the clarification process.
Our approach is designed as an AI chain that consists of five steps, each handled by a separate LLM call, to improve accuracy, efficiency, and fluency for query clarification in API recommendation.
We verify the usefulness of each unit in our AI chain, which all received high scores close to a perfect 5.
When compared to the baselines, our approach shows a significant improvement in MRR, with a maximum increase of 63.9\% higher when the query statement is covered in KG and 37.2\% when it is not. 
Ablation experiments reveal that the guidance of knowledge in the KG and the knowledge-guided pathfinding strategy are crucial for our approach's performance, resulting in a 19.0\% and 22.2\% increase in MAP, respectively.
Our approach demonstrates a way to bridge the gap between KG and LLM, effectively compensating for the strengths and weaknesses of both.
 
\end{abstract}


\section{Introduction}
\begin{figure*}
    \centering
    \includegraphics[width=0.8\textwidth]{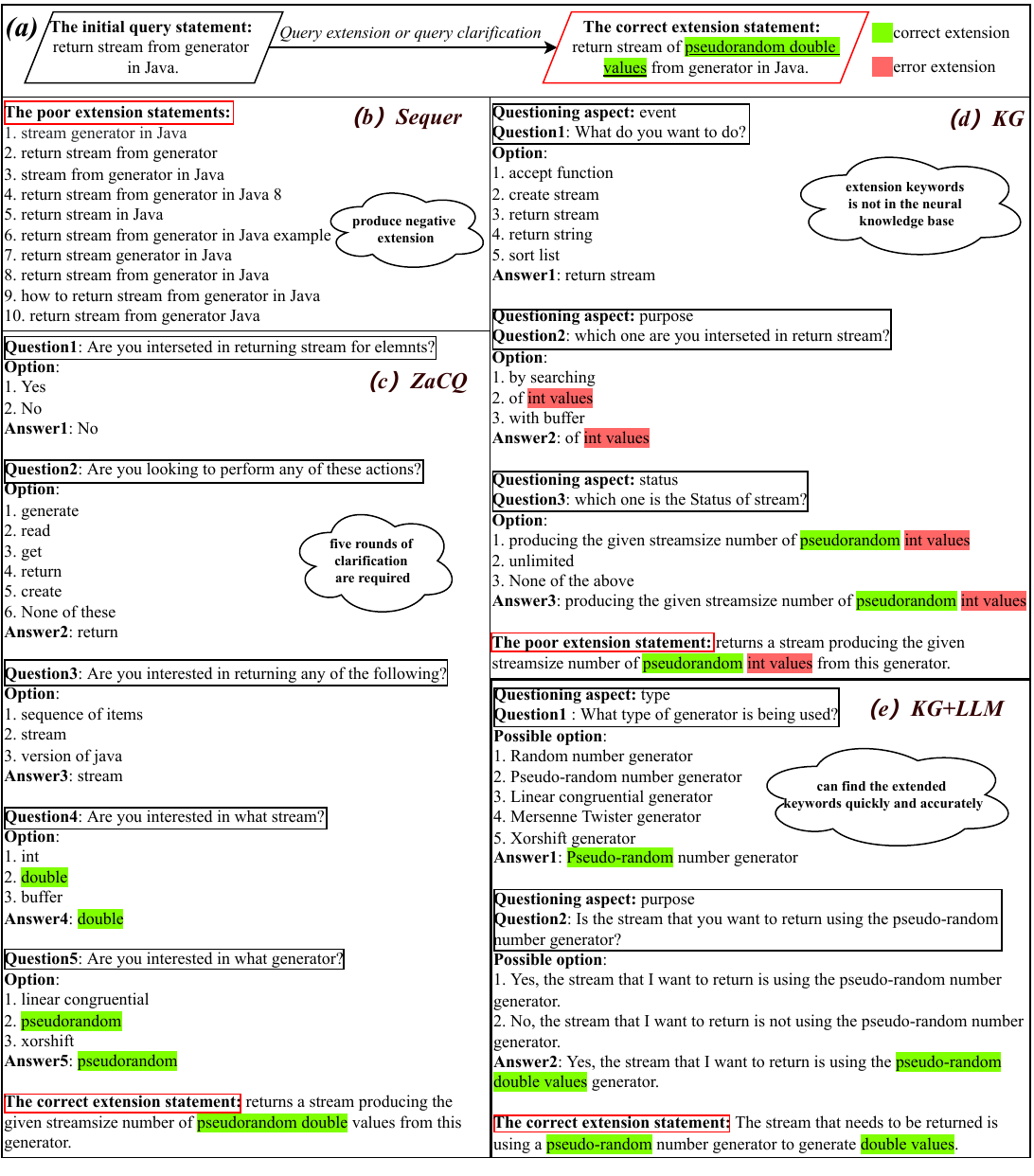}
    \vspace{-5mm}
    \caption{Existing Methods vs. Our Approach}
    \label{fig: killing example}
    \vspace{-6mm}
\end{figure*}
Application programming interface (API) recommendation is the process of finding the right API from a library based on a query statement~\cite{huang2018api, stylos2006mica, ling2019graph, bajracharya2010searching}. 
Early methods used fuzzy keyword matching to retrieve APIs that matched the literal meaning of keywords such as the API name~\cite{Thung2013AutomaticRO}, requirement description~\cite{Rahman2016RACKAA}, and tags~\cite{stylos2006mica, bajracharya2010searching}. 
However, this approach was limited as it could not capture semantic relationships between keywords. 
To address this limitation, deep learning methods were proposed to capture semantic relationships~\cite{huang2018api, ling2019graph}. 
However, these methods require query statements that contain enough keywords to reflect the user's needs.
In practice, developers often provide initial query statements with insufficient keywords to fully express their intent, which can reduce the effectiveness of deep learning methods~\cite{liu2021opportunities, zhang2017expanding}.

To address the issue of insufficient keywords in initial queries, query extension methods~\cite{cao2021automated, eberhart2022generating, huang2023answering} are proposed, such as Sequer~\cite{cao2021automated}, which retrieves query-related extension keywords from a knowledge base and modifies or adds them to the initial query statement.
However, these methods may produce negative extensions because it can be challenging to fully express the user's extension needs with only the keywords from the initial query statement, and if the correct extension keywords are not present in the knowledge base, out-of-vocabulary (OOV~\cite{Huang2022112PK}) failures can occur.
For example, Fig.~\ref{fig: killing example}-(a) shows an initial query statement ``\textit{return stream from generator in Java.}'' and a correct extension statement that aligns with user needs is  ``\textit{return stream of pseudorandom double values from generator in Java.}''. 
The the correct extension keywords in this case are ``\textit{pseudorandom double values}''. 
When using Sequer to extend the initial query statement, it produces a set of extension statements that only involve reorganizing the initial query and adding unrelated extension keywords, such as ``\textit{Java 8}'', which are entirely irrelevant to the correct extension statement, as shown in Fig.~\ref{fig: killing example}-(b).

Query clarification methods~\cite{eberhart2022generating, huang2023answering} are better than query extension methods at capturing the user's needs through interaction. 
These methods ask clarifying questions, provide options to interact with the user, and refine the query statement based on the user's response.
Deep learning-based query clarification methods, such as ZaCQ~\cite{eberhart2022generating}, are particularly useful for expressing the user's intent. 
However, ZaCQ's performance depends on the diversity and size of the training data. 
If the correct extension keywords are not present in the training data set, OOV failures may still occur.
Even if the extension keywords are present in the training dataset, ZaCQ may require multiple rounds of question clarification to accurately capture the user's intent and obtain the correct extension keywords. 
ZaCQ does not account for the relationship between each round of clarification questions, leading to inefficiency.
For example, in Fig.~\ref{fig: killing example}-(c), through five rounds of question clarification, ZaCQ obtains the correct extension statement ``\textit{returns a stream producing the given streamsize number of pseudorandom double values from this generator}''.

Huang et al. propose a Knowledge Graph (KG)-based query clarification method to improve the efficiency of query clarification methods \cite{huang2023answering}.
The KG contains a large number of APIs and their clarification information in the form of aspects (e.g., purpose, type, and event) and options.
This method searches a small-scale subgraph from KG based on an initial query statement, which contains all APIs matching the initial query statement and their corresponding clarification information. 
Based on this subgraph, an ID3 decision tree~\cite{hssina2014comparative} is constructed to provide the optimal clarification path for the user, optimizing the process of interacting with the user for clarification purposes.
However, the method is limited by the KG's scope. 
When queries fall outside the KG's scope, OOV failures can still occur, leading to irrelevant clarification questions, options or inaccurate results.
Furthermore, although clarification options (e.g., int values) may remind the user the missing information of ``double values'', the user can only select the options from the KG but cannot enter arbitrary information.
Furthermore, although clarification options (e.g., int values) may remind the user the missing information of ``double values'', the user can only select the options from the KG but cannot enter arbitrary information.
Additionally, the method relies on pre-designed templates to process the information in the KG, resulting in conversations that may feel somewhat rigid and lack fluidity.
For example, in Fig.~\ref{fig: killing example}-(d), it only takes three rounds of question clarification to obtain an extension statement, but it incorrectly extends with ``\textit{pseudorandom int values}'' instead of ``\textit{pseudorandom double values}'' due to the latter being absent in the KG. 

To address OOV failures and improve the fluency of conversations, we think of utilizing large language models (LLMs) such as GPT-3~\cite{brown2020language} as a neural knowledge base for query clarification. 
LLMs have two significant advantages. 
First, they use the entire Internet corpus (Common Crawl~\cite{luccioni2021s}) for pre-training, providing a wealth of API knowledge that overcomes OOV failures. 
Second, they possess enhanced semantic understanding and in-context learning capabilities~\cite{wan2022they, huang2022prompt, huang2022se}, which allow for more precise generation of fluent clarification queries and options.
However, LLMs also have two disadvantages. 
First, their broad knowledge base~\cite{brown2020language, nori2023capabilities} may introduce noise during clarification question and option generation, which could affect their accuracy. 
Second, LLMs exhibit a degree of randomness~\cite{lee2022coauthor, brown2020language}, which could result in limited control over the direction of clarification and reduced controllability when relying solely on LLMs for clarifications.

To address the potential noise and randomness issues associated with using LLMs, we draw on Huang et al.'s KG-based query clarification method~\cite{huang2023answering}. 
The KG provides structured API knowledge and entity relationships~\cite {huang20221+, li2018improving, liu2020generating}, resulting in reliable and comprehensive information that aids in filtering out noise from other fields.
This increases the effectiveness in generating accurate and relevant clarification questions. 
By transferring the optimal clarification path from the KG to the LLM, we improve our control over the LLM, increasing the efficiency of the clarification process and enabling the LLM to present the most appropriate clarification questions. 
This approach helps address the potential precision and randomness problems associated with solely relying on the LLM for raising clarification questions.
For instance, in Fig.~\ref{fig: killing example}-(e), the LLM, guided by the KG, accurately clarifies the user's intent in just two Question\&Answer (Q\&A) rounds, and allows the user to be inspired and enter the needed information freely (e.g., pseudo-random double values)
yielding a correct expansion statement that fully aligns with the user's intent.

Drawing on our comprehensive analysis, we propose a knowledge-guided query clarification approach for API recommendation. By combining the strengths of KG and LLM and compensating for their respective weaknesses, we achieve a ``\textit{1+1$>$2}'' effect, resulting in an optimized query clarification process. The result is an enhancement of the accuracy, efficiency, and fluency of the human-machine interaction process.

To facilitate the multi-round human-machine interaction involved in query clarification, we have designed an AI chain that breaks down our approach into several sub-steps, each handled by a separate LLM calling. Our AI chain comprises five distinct steps as follows:
\begin{itemize}
    \item \textit{Best Questioning Aspect Generation}, produce the best questioning aspect for a given query statement.
    \item \textit{Clarification Question Generation}, creates the clarification question for the query statement based on the best questioning aspect.
    \item \textit{Alternative Option Generation}, which generates alternative options in response to the clarification question.
    \item \textit{Query Extension},generates an expansion query using the historical answers.
    \item \textit{API Recommendation}, provides resulting APIs based on the expansion statement.
\end{itemize}

We conduct three experiments to evaluate the performance of our approach.
First, we verify the usefulness of each unit in the our AI chain.  The five units respectively received average scores of 4.507, 4.485, 4.455, 4.480 and 4.467, all close to a perfect 5 for usefulness. 
Second, in two scenarios, where the query statement is in or not in the KG, we compare our approach with two baselines KAHAID\cite{huang2023answering} and ZaCQ\cite{eberhart2022generating}.
For query statement in KG, our approach outperform baselines achieving 16.9\% higher MRR and 8.9\% higher MAP than KAHAID, and a significant improvement over ZaCQ with a 63.9\% higher MRR and 66.8\% higher MAP. 
For query statement not in KG, our approach achieves at least a 37.2\% higher MRR and 27.8\% higher MAP than KAHAID.
Finally, we conduct an ablation experiment to evaluate the impact of eliminating the guidance of KG knowledge and knowledge-guided pathfinding in our approach. 
The results show that utilizing KG knowledge to guide LLM helps to reduce potential noise and resulted in a 15.8\% improvement in MRR and a 19.0\% improvement in MAP for our approach. 
Additionally, the knowledge-guided pathfinding strategy is crucial for enhancing LLM's efficiency and controllability, resulting in a 19.2\% increase in MRR and a 22.2\% increase in MAP for our approach.
In summary, our approach demonstrated high precision, robustness, and controllability, accurately generating clarification questions and extending queries aligned with user intent.
 
The main contributions of this paper are as follows:
\begin{itemize}
    \item
    Our proposed approach for query clarification in API recommendation is distinct from separate KG-based or LLM-based methods. We are the first to integrate KG to guide the LLM, resulting in improved accuracy, efficiency, and fluency.
    \item
    We leverage the LLM as neural knowledge base to address OOV failures of the limited KG, and use KG to mitigate the potential noise and randomness issues associated with LLMs. 
    This opens a door to bridge the gap between KG and LLM, demonstrating how to utilize their strengths and weaknesses to effectively compensate each other.
    \item
    We design an AI chain that divides our approach to into five steps, with each step handled by a separate LLM call.
    This design enhances the robustness and controllability of our approach.  
    \item 
    A Knowledge-Guided Pathfinding Strategy is designed to transfer the optimal clarification paths in the KG to the LLM, effectively guiding the LLM to generate clarification questions along the optimal path.
    \item
    We thoroughly evaluate our approach design and find that it outperforms the baselines, irrespective of the presence of extension keywords in the KG. Our experimental results provide compelling evidence of the effectiveness of our approach.
\end{itemize}

\begin{figure*}
    \centering
    \includegraphics[width=0.9\textwidth]{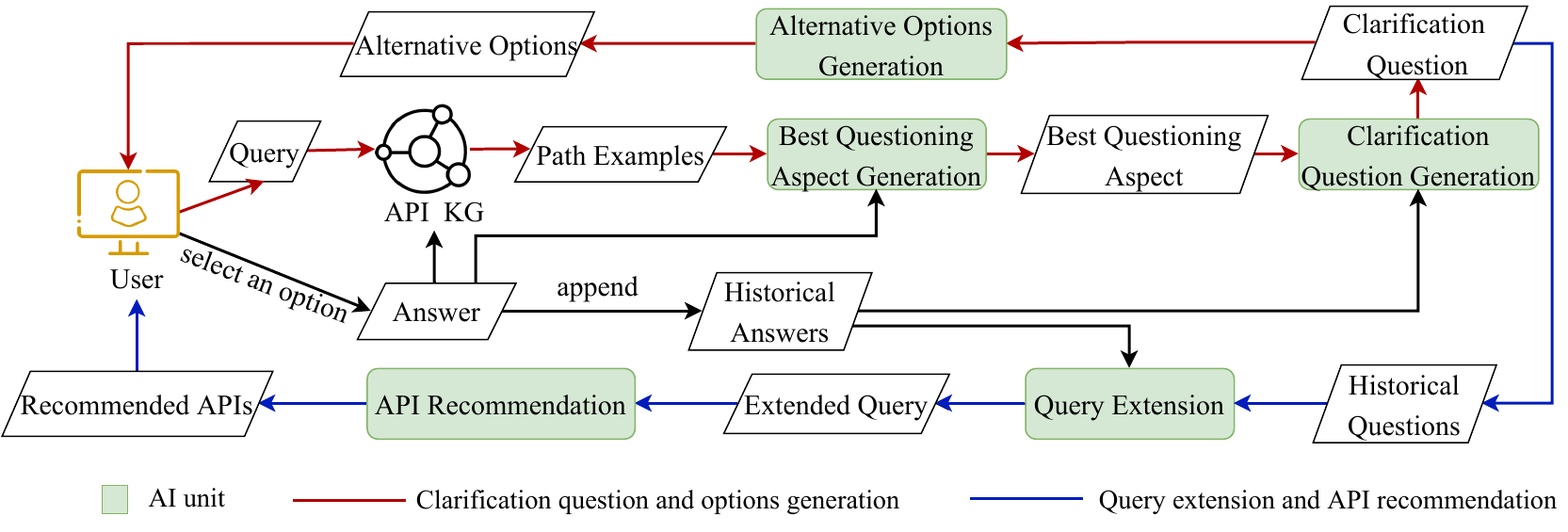}
    \vspace{-3mm}
    \caption{Overall Architecture Design}
    \label{fig: approach}
    \vspace{-2mm}
\end{figure*}
\section{The Approach}

We present KPL (KG prompt LLM), a novel knowledge-guided query clarification approach for API recommendation that combines the strengths of KG and LLM while compensating for their respective weaknesses. 
Our approach optimizes the query clarification process, enhancing the accuracy, efficiency, and fluency of the entire human-machine interaction process and improving API recommendations. 
We utilize GPT-3.5\footnote{\href{https://openai.com/blog/introducing-chatgpt-and-whisper-apis}{{https://openai.com/blog/introducing-chatgpt-and-whisper-apis}}} as the interactive LLM and the API KG~\cite{huang2023answering} as the questioning guide for the LLM. 
Moreover, our approach is not limited to GPT-3.5 and can be applied with any model that has in-context learning capability.

\subsection{Architecture Design}
\label{2.1}

Our approach consists of two key parts: ``\textit{Clarification question and options generation}'' and ``\textit{Query extension and API recommendation}'', as illustrated in Fig.~\ref{fig: approach}.

In the red line part, ``\textit{Clarification question and options generation}'', the goal is to enable the LLM to ask clarification questions and generate appropriate options.
The user first inputs a query statement, and our approach designs the knowledge-guided pathfinding strategy to retrieve the optimal clarification path knowledge from the API KG's entities and relationships.
This strategy generates path examples that guide the LLM to provide the best questioning aspect through the best questioning aspect generation unit.
Once the best questioning aspect is obtained, the LLM generates a clarification question based on the best questioning aspect and the historical answers (if any) in the clarification question generation unit.
From the generated clarification question, the alternative options generation unit can produce alternative options.

In response to the clarification question and its options, the user can select an option or provide their own answer.
The answer can be utilized in several ways. 
It can be entered into the API KG, where it becomes one of the bases for the best path retrieval. 
Alternatively, it can be entered into the best question aspect generation unit, where it becomes one of the bases for the best question aspect generation together with the path examples. 
Finally, the answer can be appended to the historical answers.
The historical answers can also be used in different ways. 
They can be entered into the clarification question generation unit, where they become one of the bases for the clarification question generation. 
Alternatively, they can be entered into the query extension unit, where they become one of the bases for the extended query generation.

In the blue line part, ``\textit{Query extension and API recommendation}'', once the clarification question is obtained, it is appended to the historical questions.
The query extension unit then extends the query statement based on the historical Q\&As. 
Finally, the API recommendation unit recommends APIs based on the extended query to the user.

\vspace{-1mm}
\subsection{Knowledge-Guided Pathfinding}
\label{strategy}
The knowledge-guided pathfinding strategy is the key to the whole approach, which transfers the optimal clarification paths in the KG to the LLM.
By learning the optimal clarification path in Knowledge Graph (KG), the accuracy and efficiency of LLM in the process of human-machine interaction can be improved.
The optimal clarification path is the shortest and most accurate route from the query statement to the relevant API. 
It includes the aspects need to be clarified and multiple rounds of clarification questions with options.

To acquire the optimal clarification paths from the KG for LLM learning, we employ the KAHAID method~\cite{huang2023answering} to construct a sample table. 
KAHAID is a KG-based query clarification method, which involves building an API KG containing a vast number of clarification information that constitutes the ingredients of optimal clarification paths.
In order to get accurate optimal clarification path to the sample table, we use KAHAID to get optimal clarification path with the queries covered in KG, i.e., 6000 API description in API KG~\cite{huang2023answering}.
These queries, when entered into KAHAID, are matched to one or more related APIs, and a path of best clarification from the query to that API is generated for each API.
We collect the optimal clarification path for all queries and add they into the best questioning path table.

Table~\ref{table: The Best Questioning Path in KG} presents a optimal clarification path collected by KAHAID~\cite{huang2023answering}, which starts with an query statement(``return stream from generator in Java'') and ends with a related API (``java.util.Random.ints''). 
The path consists of three rounds of question clarification in different aspects that ultimately discover the API.
Each round records the \textit{Aspect} that need to be clarified, the clarification questions (\textit{CQ}) used, and the \textit{Option} selected.
The \textit{Aspect} recorded in each round refers to the most significant under-specified part of the query statement that needs to be clarified. 
The \textit{CQ} records the clarification question proposed based on the \textit{Aspect}.
For example, in the first round of clarification, the \textit{Aspect} is ``event'', and then the clarification question ``What do you want to do?'' is proposed to clarify the event of the query statement.
The answer to this question, referred to as an \textit{Option}, directly affects the \textit{Aspect} of the next round.
Starting from the second round, each round's \textit{Aspect} is directly related to the previous round's \textit{Option}.
For example, after selecting the \textit{Option} ``return stream'' in the first round, the purpose of ``return stream'' is selected as the \textit{Aspect} of the second round, and a clarification question is generated immediately to further clarify its purpose.

\begin{table*}[h]
\centering
\caption{The Best Questioning Path in KG}
\vspace{-1mm}
\label{table: The Best Questioning Path in KG}
\resizebox{0.95\textwidth}{!}{
\begin{tabular}{|c|c|c|c|c|c|c|}
\hline
Query                                                                                                  & Round  & Aspect  & CQ                                                                                         & Option                                                                                                          & \begin{tabular}[c]{@{}c@{}}Next Round\\ Aspect\end{tabular} & API                                                                                     \\ \hline
\multirow{6}{*}{\begin{tabular}[c]{@{}c@{}}return \\ stream\\ from\\ generator\\ in Java\end{tabular}} & Round1 & event   & \begin{tabular}[c]{@{}c@{}}What do you what to do?\end{tabular}                          & \begin{tabular}[c]{@{}c@{}}return stream\end{tabular}                                                         & purpose                                                     & \multirow{6}{*}{\begin{tabular}[c]{@{}c@{}}java.\\ util.\\ Random.\\ ints\end{tabular}} \\ \cline{2-6}
                                                                                                       & Round2 & purpose & \begin{tabular}[c]{@{}c@{}}Which one are you interested\\ in return stream?\end{tabular} & of int value                                                                                                    & status                                                      &                                                                                         \\ \cline{2-6}
                                                                                                       & Round3 & status  & \begin{tabular}[c]{@{}c@{}}Which one is the \\status of stream?\end{tabular}             & \begin{tabular}[c]{@{}c@{}}producing the given\\ streamsize number of\\ pseudorandom int value\end{tabular} & -                                                           &                                                                                         \\ \hline
\end{tabular}}
\vspace{-4mm}
\end{table*}

To find optimal clarification path, we design a knowledge-guided pathfinding strategy that extracts path knowledge from Table~\ref{table: The Best Questioning Path in KG}, which corresponds to API KG in Fig.~\ref{fig: approach}. 
The initial query statement entered by the user and the previous round answer (set to ``\textit{None}'' by default in the first round) are retrieved from Table~\ref{table: The Best Questioning Path in KG} to match similar paths. 
First, the query statement is matched with the ``\textit{Query}'' in Table~\ref{table: The Best Questioning Path in KG} by similarity, and the top 10\% data with the highest similarity are selected. 
Next, the ``\textit{Option}'' in this 10\% of the path data is ranked by similarity with the previous round answer.
Finally, we extract five path examples with different aspects as path knowledge to guide the LLM based on the sorted path data.
We perform similarity ranking twice to ensure the path retrieved from the API KG is similar to the query statement entered by the user.
The selection of five path examples with different aspects ensures the diversity of path knowledge.

\subsection{Prompt Design for AI-Units}
Our approach is implemented as an AI chain, which involves five distinct AI units corresponding to five steps.
In this section, we discuss how to design natural language prompts that direct the LLM to perform specific AI functionalities.

\begin{figure*}
    \centering
    \includegraphics[width=0.9\textwidth]{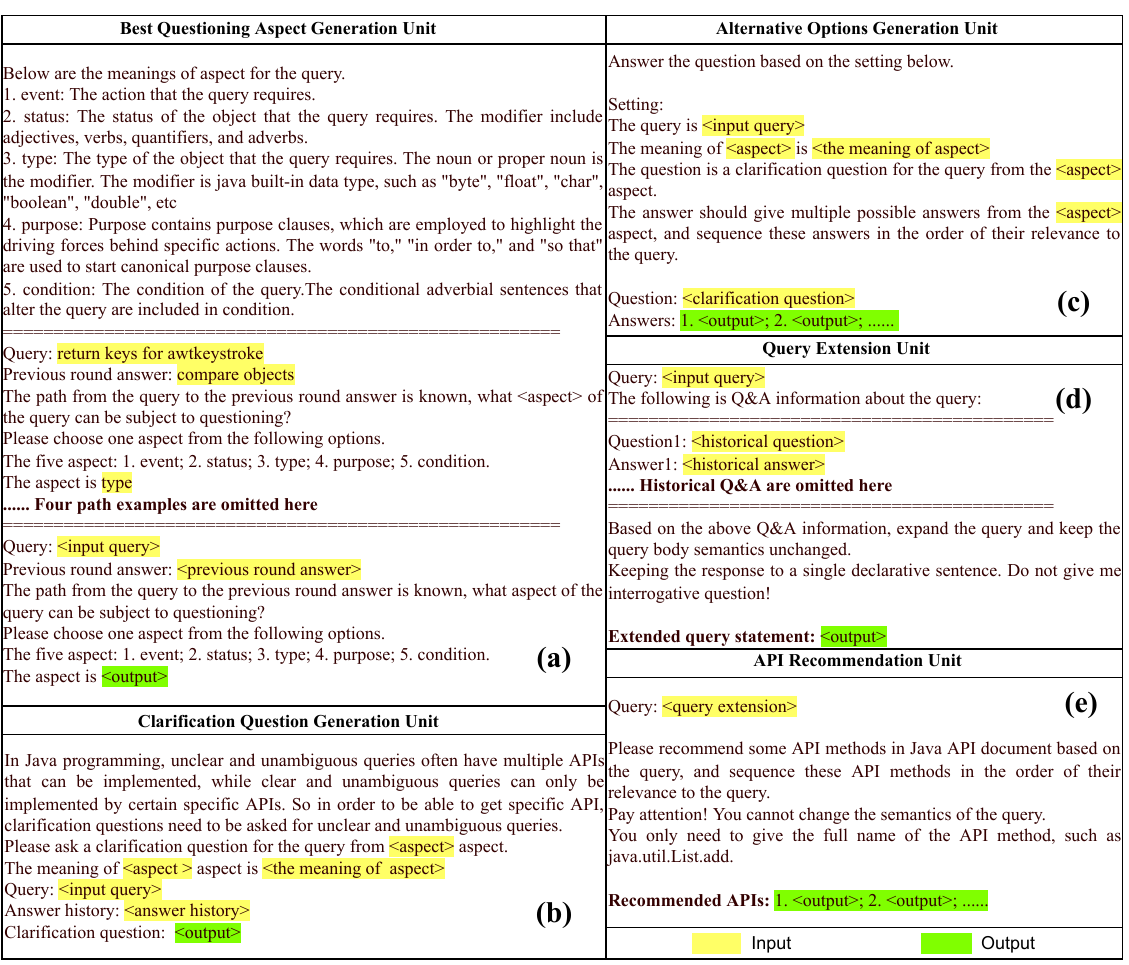}
    \vspace{-3mm}
    \caption{Prompt Design for Five AI Unit}
    \label{fig: Five AI Unit}
    \vspace{-5mm}
\end{figure*}
\subsubsection{Best Questioning Aspect Generation Unit}
To enable the LLM to ask efficient, robust, and controlled clarification questions along the optimal clarification path, determining the best questioning aspect is essential. 
We introduce an AI unit that uses the initial query statement and the previous round answer to determine the best questioning aspect. 
The contents of the prompt are illustrated in Fig.~\ref{fig: Five AI Unit}-(a), which includes the meaning of the five aspects, the five path examples retrieved from the API KG, and the input in the form of examples. 
Importantly, the data in the KG serves only as a prompt example for the LLM, rather than directly impacting its decision-making.
Its role is to guide the LLM in making informed decisions based on the given examples.
Each example consists of two inputs, the query statement and the previous round answer, and their output, the best questioning aspect.

\subsubsection{Clarify Question Generation Unit}
This AI unit is responsible for generating clarification questions based on the best questioning aspect obtained in the previous step. This task is described in Fig.~\ref{fig: Five AI Unit}-(b) as ``\textit{In Java programming, ...}'' followed by ``\textit{Please ask a clarification ...}'' and the current query and known information. 
Additionally, to ensure that the LLM understands the meaning of the selected aspect, the task description includes ``\textit{The meaning of $<$aspect$>$}''.

\subsubsection{Alternative Options Generation Unit}
This AI unit is responsible for generating alternative options based on the clarification question asked by the previous AI unit. 
The prompt for this unit is shown in Fig.~\ref{fig: Five AI Unit}-(c), which includes a task description of ``\textit{Answer the question based on the setting below}'' and some specific settings for the task. 
The input for this unit is the clarification question generated in the current round. 
The output of this unit is a ranked list of alternative options for the clarification question. 
By providing the unit with a clarification question and its specific settings, the LLM can learn and generate appropriate alternative options for the clarification question.

\subsubsection{Query Extension Unit}
This AI unit is responsible for generating an extended query statement based on the user's needs after the clarification process. 
The prompt for this unit is shown in Fig.~\ref{fig: Five AI Unit}-(d), which includes the input query statement, historical Q\&A, and a task description ``\textit{Based on the above Q\&A ...}''.
By providing the historical Q\&A information, the unit can identify the user's missing needs and extend the query statement accordingly.

\subsubsection{API Recommendation Unit}
This unit enables the user to refer to the recommended APIs after each clarification and decide whether to continue with the clarification process or not, thus improving the flexibility of interaction.
The prompts for this unit are displayed in Fig.~\ref{fig: Five AI Unit}-(e) and consist of a task description, ``\textit{Please recommend some API methods ...}'', and some generation restrictions, ``\textit{Pay attention! You cannot ...}''.
The input for this unit is the extended query statement generated in the previous unit, and the output is a list of multiple possible APIs for the extended query statement, based on the task description and generation restrictions provided.

\subsection{Running Example}
\begin{figure}
    \centering
    \vspace{-9mm}
    \includegraphics[width=0.42\textwidth]{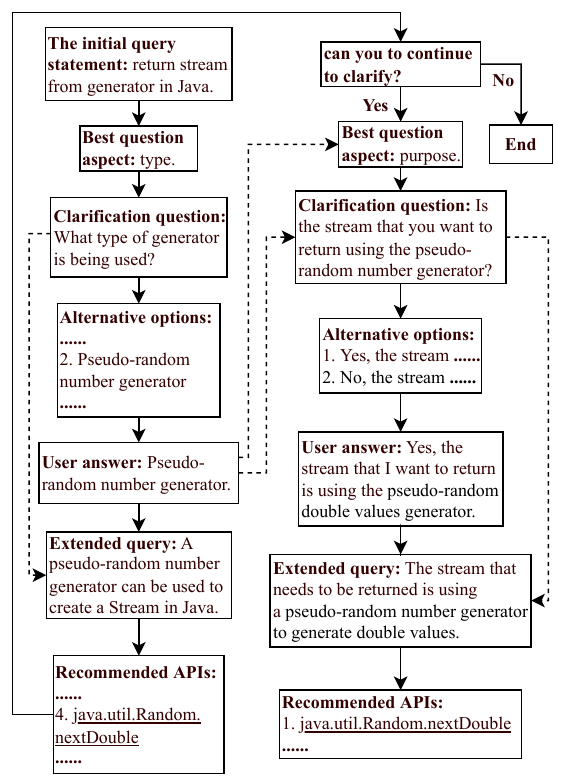}
    \vspace{-3mm}
    \caption{Running Example}
    \label{fig: Running Example}
    \vspace{-4mm}
\end{figure}
We provide a running example to demonstrate how the AI units work together and how users can interact with the LLM, as shown in Fig.~\ref{fig: Running Example}. The solid line represents the flow of the method, and the dashed line represents the data flow.

The first step is to input the initial query statement and the previous round answer into the \textit{Best Questioning Aspect Generation Unit}, which generates the best questioning aspect, in this case, is ``\textit{type}''. 
If it is the first round of questioning, the previous round answer is assumed to be ``\textit{None}''.

Next, the \textit{Clarification Question Generation Unit} generates the clarification question for the initial query statement based on the best questioning aspect and historical answers. Then, the \textit{Alternative Options Generation Unit} generates some alternative options based on the clarification question.

Upon receiving a clarification question and alternative options, the user provides an answer that references these alternatives. The user's previous questions and answers are then input into the \textit{Query Extension Unit}, which generates an extended query statement based on this historical data.

Subsequently, the extended query statement is fed into the \textit{API Recommendation Unit}, which recommends the corresponding APIs based on the extended query statement and sorts them by relevance. 
In this case, the best API is ``\textit{java.util.Random.nextDouble}''. 
However, in the first round, it isn't very relevant. 
The user continues to proceed to the next round of clarification dialogue.

Repeat the steps similar to the first round to get a new round of recommended APIs. 
In this round, the best API comes first and is most relevant. 
Finally, the user gets the best API and ends the clarification dialogue.

\section{Experiment Setting}
\vspace{-2mm}
This section presents research questions to evaluate the performance of our approach, along with experimental setup including data preparation, baselines, and evaluation metrics.

\subsection{Research Questions}\label{RQ}
We formulate three research questions to evaluate the performance of KPL in query clarification, as follows:
\begin{itemize}
    \item RQ1: What is the effectiveness of each AI unit?
    \item RQ2: How well does KPL perform in extending query statements across two scenarios, namely, query statement covered in API KG and those not covered in API KG?
    \item RQ3: How effective is the knowledge-guided pathfinding strategy employed in our approach?
\end{itemize}

\subsection{Data Preparation}
We have designed two test datasets, KG Query Statement (KGQS) and non-KG Query Statement (non-KGQS), to comprehensively evaluate the performance of KPL in two different scenarios: query statements covered and not covered in the API KG.
This comparison enables us to confirm that the KG serves as a prompt example for LLMs, effectively guiding their decision-making, without directly substituting for LLMs in making decisions.

1) \textbf{KG Query Statement (KGQS)}: For fairness, this dataset contains 60 query statements randomly selected from KAHAID's API KG, each paired with its corresponding ground-truth API~\cite{huang2023answering}.
As mentioned in Section~\ref{2.1}, we use 6,000 API descriptions from the API KG to obtain an example of the best path for guiding the LLM. 
To satisfy the scenario of ``query statement covered in KG'', we randomly select 60 API descriptions from the 6,000 API descriptions as query statements and use their related APIs in the API KG as the ground-truth APIs.

2) \textbf{non-KG Query Statement (non-KGQS)}: Also for the sake of fairness, this dataset is the evaluation dataset of KAHAID, a state-of-the-art approach, which contains 60 queries each paired with multiple ground-truth APIs.
We reuse this high quality dataset of KAHAID, which are manually collected from the Stack Overflow using strict criteria, which could be found in KAHAID's paper~\cite{huang2023answering}.
It's worth noting that none of the query statements in this dataset are covered in the API KG.

\vspace{-1mm}
\subsection{Baselines}
To evaluate KPL's performance in query clarification, we compare it with two existing methods, ZaCQ~\cite{eberhart2022generating} (a deep learning-based method) and KAHAID~\cite{huang2023answering} (a KG-based method) on two datasets (KGQS and non-KGQS).

Furthermore, we conduct ablation experiments on KPL in RQ3 to explain its working principle. 
We design two variants:
\begin{itemize}
    \item
    KPL$_{w/oK}$, which does not use KG to guide the LLM on the best questioning aspect.
    \item 
    KPL$_{w/oKPS}$, which does not use the knowledge-guided pathfinding strategy for the best questioning aspect but still has KG to guide the LLM to ask clarification questions.
    Specifically, KPL$_{w/oKPS}$ only retrieves the query statement to obtain the best questioning aspect, but KPL first retrieves the query statement and then retrieves the previous round answer to obtain the best questioning aspect, see Section~\ref{strategy} for details.
\end{itemize}
We compare KPL$_{w/oK}$ with KPL to assess the effectiveness of KG guidance, and KPL$_{w/oKPS}$ with KPL to assess the effectiveness of the knowledge-guided pathfinding strategy.

\vspace{-1mm}
\subsection{Evaluation Metrics}
\label{metrics}
In RQ1, we use questionnaires to assess the quality of each AI unit.
We enlist three master's students with over three years of coding experience and API knowledge, who are not involved in our approach design and implementation, to score the questionnaire and ensure its rationality.
They rate our tool based on various factors~\cite{blackwell2000cognitive}, including:

\begin{itemize}
\item
Intelligibility refers to whether the questions and answers generated by the tool are clear and easy to understand, and whether the language (terminology or expression) is straightforward and understandable.
\item
Diversity refers to whether the tool generates a variety of questions and answers, or if it produces numerous repeated or similar questions and answers, of if it generate different types and styles of questions and answers. 
\item
Relevance refers to whether the questions and responses generated by the tool are relevant to the input content, can summarize the input's theme or key points, and address the user's actual needs. 
\item
Trustworthiness refers to whether the questions and answers produced by the tool are credible, conform to domain specifications, and can provide professional-level questions and answers.
\item
Language fluency refers to whether the conversation generated by the tool are natural and smooth, whether the language used is grammatically correct and semantically sound, and whether there are any ambiguities. 
\item
Practicality refers to whether the tool can solve the users' actual needs, whether it has practical application value, and whether it is easy to use.
\end{itemize}

The questionnaire uses a five-point scale to score each factor, with higher scores indicating a more positive attitude towards the tool's performance, and lower scores indicating a more negative attitude. 
A score of 5 indicates strong conformity, while a score of 1 indicates very poor conformity, and a score of 3 indicates neutrality. 
While all evaluation aspects are important, relevance and credibility are the most crucial factors. 
We assign weights for each factor, with 15\% for intelligibility, 15\% for diversity, 20\% for relevance, 20\% for trustworthiness, 15\% for language fluency, and 15\% for practicality.
Note that here we assign higher weights to relevance and trustworthiness. Relevance signifies the correlation between the output of the LLM and the requirements. Trustworthiness is crucial for ensuring the reliability of the information.

In RQ2 and RQ3, we evaluate the performance of query clarification methods and KPL using metrics such as Mean Reciprocal Rank (MRR)~\cite{Radev2002EvaluatingWQ}, Mean Average Precision (MAP)~\cite{Sanderson2010ChristopherDM}, Precision, and Recall.
MRR and MAP have commonly used metrics for evaluating information retrieval systems. MRR measures the level of effort required to find the first accurate answer in the list of recommendations, while MAP takes into account the ranks of all correct answers. In addition, we will also use Precision and Recall to evaluate the performance of the API recommendation approaches.
Precision measures the proportion of recommended APIs that are ground-truth APIs among all recommended APIs.
Recall represents the proportion of ground-truth APIs that are recommended by our approach and baseline.

\section{Experiment Analysis}
This section delves into three research questions to evaluate and discuss the performance of our approach.

\vspace{-1mm}
\subsection{RQ1: What is the effectiveness of each AI unit?}

\begin{table}[]
\centering
\caption{The performance of each AI unit}
\vspace{-2mm}
    \footnotesize
        Note: These results are the total score obtained by multiplying each of the six evaluation factors by their respective weights. 
\label{table: The performance of each AI unit}
\resizebox{0.48\textwidth}{!}{
\begin{tabular}{|c|c|c|c|c|}
\hline
AI unit                            & Grader1 & Grader2 & Grader3 & \begin{tabular}[c]{@{}c@{}}Average\\ Rating\end{tabular} \\ \hline
\begin{tabular}[c]{@{}c@{}}Best Questioning\\ Aspect Generation\end{tabular}  &4.455 &4.485 &4.565 & 4.501   \\ \hline
\begin{tabular}[c]{@{}c@{}}Clarification Question\\ Generation\end{tabular}  & 4.485 & 4.495 &4.475 &4.485 \\ \hline
\begin{tabular}[c]{@{}c@{}}Alternative\\ Options Generation\end{tabular} & 4.440  &4.465  &4.460  &4.455  \\ \hline
\begin{tabular}[c]{@{}c@{}}Query\\ Extension\end{tabular} 
                       & 4.475 &4.505   &4.460   &4.480  \\ \hline

\begin{tabular}[c]{@{}c@{}}API\\ Recommendation\end{tabular}       
        & 4.480 &4.470    & 4.450 &4.467  \\ \hline
\end{tabular}}
\vspace{-7mm}
\end{table}

\subsubsection{Motivation}
Our approach involves user interaction, making it impractical to integrate the query clarification task into a single prompt for LLM's in-context learning. 
So we design an AI chain comprising five AI units, each handled by a separate LLM call. 
In this RQ, we investigate the effectiveness of each AI unit in fulfilling its respective responsibilities.

\subsubsection{Methodology}
We apply KPL to the non-KGQS dataset and collect the intermediate results produced by each AI unit.
We assign 10 test cases to each participant, and participants will have 2 minutes to fill out the questionnaire based on the criteria outlined in Section~\ref{metrics} after each test. We collect the questionnaires filled out by all participants and calculate the average score.
The results are presented in Table~\ref{table: The performance of each AI unit}, and more detailed metric information can be found in Section~\ref{metrics}.

\subsubsection{Result Analysis}
We use a five-point scale as our rating system.
Table~\ref{table: The performance of each AI unit} shows that the average rating in each AI unit we designed is high, indicating these AI units' effectiveness.
The first AI unit, \textit{Best Questioning Aspect Generation}, achieves an average rating of 4.501, demonstrating its ability to generate the best questioning aspect due to the effective guidance provided by the optimal path from API KG.
For example, in Fig.~\ref{fig: Running Example}, the first round of questioning aspect proposed by this unit to the initial query statement is ``\textit{type}''.
However, in Fig.~\ref{fig: killing example}-(d), the first round of questioning aspect of KG is ``\textit{event}''. 
This is because LLM has learned better path knowledge from the examples in the prompt. 

The second unit, \textit{Clarification Question Generation}, achieves an average rating of 4.485, indicating its ability to generate clarification questions that conform to the best questioning aspect.
For instance, in Fig.~\ref{fig: Running Example}, the first round of questioning generated by this unit aligns with the best questioning aspect of ``\textit{type}'' and results in the question ``\textit{What type of generator is being used?}''.

The third unit, \textit{Alternative Options Generation}, achieves an average rating of 4.455, suggesting that it is effective in generating alternative options that are relevant to the clarification question and the initial query statement. For example, in Fig.~\ref{fig: Running Example}, this unit provides five different corresponding options for the clarification question generated by the second unit, according to the initial query statement.

The fourth unit, \textit{Query Extension}, achieved an average rating of 4.480, indicating that it is capable of generating natural, comprehensive query statements that capture the user's needs. This is achieved by leveraging LLM's powerful semantic understanding and in-context learning capabilities. For example, in Fig.~\ref{fig: Running Example}, the extended query statement ``\textit{The stream...}'' contains the initial query statement and the user's clarified needs after two rounds of question clarification.

The fifth AI unit, \textit{API Recommendation}, achieved an average rating of 4.467. It also utilizes LLM's powerful semantic understanding and in-context learning capabilities to generate corresponding APIs based on the extended query statement and rank them by relevance. For example, in Fig.~\ref{fig: Running Example}, seven APIs are recommended based on the extended query statement ``\textit{the stream...}'', with the most relevant API, \textit{java.util.Random.nextDouble}'', ranked first.

In order to ensure consistency, we tested the kappa of each person's questionnaire score, and each person's kappa for each AI unit was higher than 0.6213.

\vspace{1mm}
\noindent\fbox{
\begin{minipage}{8.6cm} 
\textit{The high average rating of the AI units confirms the effectiveness of our prompt design and composition in connecting the AI units to accomplish higher-level tasks.}

\end{minipage}}

\vspace{-1mm}
\subsection{RQ2: How well does KPL perform in extending query statements across two scenarios, namely, query statement covered in API KG and those not covered in API KG?}

\begin{table}[]
\centering

\caption{The Results of Baselines VS. Our Approach}
\vspace{-2mm}
\label{table: The Results of Baselines VS. Our Approach}
\resizebox{0.48\textwidth}{!}{
\begin{tabular}{|c|c|c|c|c|c|}
\hline
Dataset                                                                 & Metrics                    & Approaches                                               & Round 1        & Round 2        & Round 3        \\ \hline
\multirow{12}{*}{KGQS}                                               & \multirow{3}{*}{MRR}       & \begin{tabular}[c]{@{}c@{}}KPL\end{tabular} & \textbf{0.883} & \textbf{0.916} & \textbf{0.941} \\ \cline{3-6} 
                                                                        &                            & KAHAID                                                   & 0.607          & 0.756          & 0.805          \\ \cline{3-6} 
                                                                        &                            & ZaCQ                                                     & 0.524          & 0.542          & 0.574          \\ \cline{2-6} 
                                                                        & \multirow{3}{*}{MAP}       & \begin{tabular}[c]{@{}c@{}}KPL\end{tabular} & \textbf{0.830} & \textbf{0.877} & \textbf{0.904} \\ \cline{3-6} 
                                                                        &                            & KAHAID                                                   & 0.665          & 0.791          & 0.834          \\ \cline{3-6} 
                                                                        &                            & ZaCQ                                                     & 0.473          & 0.521          & 0.542          \\ \cline{2-6} 
                                                                        & \multirow{3}{*}{Precision} & \begin{tabular}[c]{@{}c@{}}KPL\end{tabular} & \textbf{0.737} & \textbf{0.813} & \textbf{0.847} \\ \cline{3-6} 
                                                                        &                            & KAHAID                                                   & 0.682          & 0.756          & 0.807          \\ \cline{3-6} 
                                                                        &                            & ZaCQ                                                     & 0.467          & 0.566          & 0.566          \\ \cline{2-6} 
                                                                        & \multirow{3}{*}{Recall}    & \begin{tabular}[c]{@{}c@{}}KPL\end{tabular} & \textbf{0.864} & \textbf{0.906} & \textbf{0.925} \\ \cline{3-6} 
                                                                        &                            & KAHAID                                                   & 0.745          & 0.789          & 0.825          \\ \cline{3-6} 
                                                                        &                            & ZaCQ                                                     & 0.583          & 0.578          & 0.588          \\ \hline
\multirow{8}{*}{\begin{tabular}[c]{@{}c@{}}non-KGQS\end{tabular}} & \multirow{2}{*}{MRR}       & \begin{tabular}[c]{@{}c@{}}KPL\end{tabular} & \textbf{0.828} & \textbf{0.869} & \textbf{0.900} \\ \cline{3-6} 
                                                                        &                            & KAHAID                                                   & 0.506          & 0.619          & 0.656          \\ \cline{2-6} 
                                                                        & \multirow{2}{*}{MAP}       & \begin{tabular}[c]{@{}c@{}}KPL\end{tabular} & \textbf{0.830} & \textbf{0.864} & \textbf{0.887} \\ \cline{3-6} 
                                                                        &                            & KAHAID                                                   & 0.508          & 0.641          & 0.694          \\ \cline{2-6} 
                                                                        & \multirow{2}{*}{Precision} & \begin{tabular}[c]{@{}c@{}}KPL\end{tabular} & \textbf{0.790} & \textbf{0.833} & \textbf{0.876} \\ \cline{3-6} 
                                                                        &                            & KAHAID                                                   & 0.604          & 0.667          & 0.697          \\ \cline{2-6} 
                                                                        & \multirow{2}{*}{Recall}    & \begin{tabular}[c]{@{}c@{}}KPL\end{tabular} & \textbf{0.858} & \textbf{0.881} & \textbf{0.891} \\ \cline{3-6} 
                                                                        &                            & KAHAID                                                   & 0.601          & 0.633          & 0.645          \\ \hline
\end{tabular}}
\vspace{-2mm}
\end{table}

\subsubsection{Motivation}
In this RQ, we compare our method with two baseline methods, ZaCQ~\cite{eberhart2022generating}  and KAHAID~\cite{huang2023answering}, to evaluate whether our approach outperforms them in scenarios where the correct extension statement is absent or present in the KG. Our objective is to demonstrate that our approach can effectively address OOV failures by showing its superiority over the baselines when the correct extension keywords are not present in the KG.

\subsubsection{Methodology}
We conduct experiments on two datasets, KGQS and non-KG, to compare our method with baseline methods. 
On the KGQS dataset, we evaluate three methods, ZaCQ, KAHAID, and KPL. 
On the non-KGQS dataset, we evaluate two methods, KAHAID and KPL, to demonstrate our approach's ability to overcome OOV failures. 
Note that here, whether it's data from the KG or not, it won't affect ZaCQ, so there's no need to evaluate ZaCQ again in the non-KGQS dataset.
The experiment results are presented in Table~\ref{table: The Results of Baselines VS. Our Approach}, and more detailed metric information can be found in Section~\ref{metrics}.

\subsubsection{Result Analysis}
Our method's superiority over baselines is demonstrated on both KGQS and non-KGQS datasets, as shown in Table~\ref{table: The Results of Baselines VS. Our Approach}. 
On the KGQS dataset, our method outperforms both baseline methods in all metrics, with particularly high values in the first round of clarification. 
For example, our MRR value in the first round of clarification is 45.5\% and 68.5\% higher than KAHAID and ZaCQ, respectively. 
As clarification rounds increase, our method and baselines show consistent increases in metric values, with our method consistently outperforming the baselines. 
For example, in the third round of clarification, our method's precision is 5.6\% and 49.6\% higher than the two baseline methods, respectively. 
The results indicate that transferring the optimal clarification path from the KG to the LLM enhances the fluidity and controllability of the conversation, enabling us to identify the user's intent efficiently and recommend API. 
For example, as shown in Fig.~\ref{fig: killing example}-(e), our approach requires only two rounds of clarification to get the correct extension statement.

On the non-KGQS dataset, our approach maintains high metric values compared to the KGQS dataset, while KAHAID's performance drops significantly. For instance, our approach's MRR value only decreases by 6.6\% in the first round of clarification, while KAHAID drops by 16.6\%. 
As the number of clarification rounds increases, KAHAID's metric values show little improvement, such as the maximum increase in MAP value, which only goes from 50.8\% to 69.4\%, indicating its inability to address OOV failures. 
However, our approach successfully handles OOV failures by utilizing the LLM as a neural knowledge base to obtain the correct extension statement even when it is not present in the KG. 
As shown in Fig~\ref{fig: killing example}-(d) and Fig.~\ref{fig: killing example}-(e), KAHAID cannot provide the correct query extension statement due to OOV failures. 
However, our approach utilizes the LLM as a neural knowledge base, allowing us to successfully obtain the correct extension statement even when it is not present in the KG. 
The slight decrease in our approach's performance on the non-KGQS dataset is due to the absence of the correct extension statement in the KG, resulting in the KG being unable to find the optimal clarification path. However, even with a suboptimal path, our approach can still perform well, demonstrating its effectiveness in addressing OOV failures. 
Despite the better performance of our approach, we still cannot guarantee absolute accuracy. This could be due to the fact that there are only five aspects of guidance in the knowledge graph, and the level of detail they provide to the language model is insufficient. On the other hand, this is also related to the capabilities of the language model itself. The language model has randomness and cannot ensure consistent outputs.

\vspace{1mm}
\noindent\fbox{
\begin{minipage}{8.6cm} 
\textit{Our approach surpasses baseline methods on both KGQS and non-KGQS datasets, showing its efficiency in addressing OOV failures in query clarification and identifying user intent to recommend necessary APIs.
}
\end{minipage}}

\subsection{RQ3:How effective is the knowledge-guided pathfinding strategy employed in our approach?}

\begin{table}[]
\centering
\caption{Ablation Results of KPL Variants}
\vspace{-2mm}
\resizebox{0.45\textwidth}{!}{
\label{table: Ablation Results of KPL Variants}

\begin{tabular}{|c|c|c|c|c|}
\hline
Metrics                    & Approaches    &Round 1     &Round 2     &Round 3     \\ \hline
\multirow{4}{*}{MRR}       & \begin{tabular}[c]{@{}c@{}}KPL\end{tabular} & \textbf{0.828} & \textbf{0.869} & \textbf{0.900} \\ \cline{2-5} 
                           & \begin{tabular}[c]{@{}c@{}}KPL$_{w/oKPS}$\end{tabular}        & 0.683 & 0.708 & 0.755 \\ \cline{2-5} 
                           & \begin{tabular}[c]{@{}c@{}}KPL$_{w/oK}$\end{tabular}          & 0.716 & 0.744 & 0.777 \\ \hline
\multirow{4}{*}{MAP}       & \begin{tabular}[c]{@{}c@{}}KPL\end{tabular} & \textbf{0.830} & \textbf{0.864} & \textbf{0.887}  \\ \cline{2-5} 
                           & \begin{tabular}[c]{@{}c@{}}KPL$_{w/oKPS}$\end{tabular}        & 0.669 & 0.690 & 0.726 \\ \cline{2-5} 
                           & \begin{tabular}[c]{@{}c@{}}KPL$_{w/oK}$\end{tabular}          & 0.690 & 0.715 & 0.745 \\ \hline
\multirow{4}{*}{Precision} & \begin{tabular}[c]{@{}c@{}}KPL\end{tabular} & \textbf{0.790} & \textbf{0.833} & \textbf{0.876} \\ \cline{2-5} 
                           & \begin{tabular}[c]{@{}c@{}}KPL$_{w/oKPS}$\end{tabular}        & 0.516 & 0.526 & 0.543 \\ \cline{2-5} 
                           & \begin{tabular}[c]{@{}c@{}}KPL$_{w/oK}$\end{tabular}          & 0.523 & 0.540 & 0.573 \\ \hline
\multirow{4}{*}{Recall}    & \begin{tabular}[c]{@{}c@{}}KPL\end{tabular} & \textbf{0.858} & \textbf{0.881} & \textbf{0.891} \\ \cline{2-5} 
                           & \begin{tabular}[c]{@{}c@{}}KPL$_{w/oKPS}$\end{tabular}        & 0.624 & 0.635 & 0.655 \\ \cline{2-5} 
                           & \begin{tabular}[c]{@{}c@{}}KPL$_{w/oK}$\end{tabular}          & 0.629 & 0.654 & 0.684 \\ \hline
\end{tabular}}
\vspace{-6mm}
\end{table}

\subsubsection{Motivation}
In this RQ, we investigate whether utilizing KG knowledge to guide LLM can reduce potential noise, and whether implementing a knowledge-guided pathfinding strategy to transfer the optimal clarification path from KG to LLM can enhance LLM's efficiency and controllability.

\subsubsection{Methodology}
We set up two approach variants (i.e., KPL$_{w/oK}$ and KPL$_{w/oKPS}$).
We run KPL$_{w/oK}$ and KPL$_{w/oKPS}$ on non-KGQS dataset and report the results in Table~\ref{table: Ablation Results of KPL Variants}.
The same method as RQ2 is employed to test these variants and calculate metric values.

\subsubsection{Result Analysis}
Table~\ref{table: Ablation Results of KPL Variants} presents the experimental results. 
KPL achieves significantly higher metrics than the other two variants. For example, in the first round of clarification, KPL's MRR value is 21.2\% and 15.6\% higher than KPL${w/oKPS}$ (without knowledge-guided pathfinding strategy) and KPL${w/oK}$ (without KG), respectively. 

The emergence of this phenomenon is due to the lack of guidance from KG and strategies for KG extraction. While LLM can indeed generate questions that need clarification, the absence of Knowledge Graph-related guidance introduces more randomness into the questioning process. This randomness results in a higher likelihood of generating incoherent questions. These findings underscore the significance of incorporating Knowledge Graph-guided assistance and strategically extracting information from Knowledge Graphs. The presence of knowledge in Knowledge Graphs, coupled with well-established pathways created through knowledge-guided traversal methods, not only plays a critical role in understanding user intent but also proves crucial in providing suggestions for necessary API.
Furthermore, although the improvement brought about by our approach is only between 15\% to 22\%, this enhancement enables our method to perform better in practical scenarios. For instance, our approach is capable of refining requirements through two rounds of clarifying questions and answers, whereas the LLM method requires more rounds and might even clarify haphazardly.

Also, KPL${w/oK}$ outperforms KPL${w/oKPS}$. 
For example, in the second round of clarification, KPL${w/oK}$'s recall value is 3.0\% higher than KPL${w/oKPS}$. 
This is because even without KG guidance, LLM can still identify the correct clarification path.
However, using only KG guidance without the knowledge-guided pathfinding strategy may result in an incorrect path extracted from the KG, leading LLM to clarify in the wrong direction. For example, for the initial query statement ``\textit{return stream from generator in Java.}'', without KG guidance, LLM can still identify the correct aspect (i.e., ``type'') as the next clarification aspect. 
But in the presence of KG guidance without the knowledge-guided pathfinding strategy, the transferred clarification path from the KG is incorrect, leading LLM to clarify in the direction of ``condition''. 

\vspace{1mm}
\noindent\fbox{
\begin{minipage}{8.6cm} 
\textit{While LLM may sometimes identify the correct clarification path without KG guidance, it can still make errors due to noise. 
Our results highlight the crucial role of combining the knowledge-guided pathfinding strategy and KG guidance in achieving optimal performance.
}
\end{minipage}}

\section{Discussion}
There are potential internal and external threats to the validity of our approach. 
Additionally, we discussed how to bridge the gap between KG and LLM.

\vspace{-1mm}
\subsection{Internal threat to the validity} 
The rating of human experience is an internal threat to the reliability of our results.
In order to mitigate this issue, we invite three master students who possessed three years of programming experience to provide experience rating through a carefully crafted questionnaire. 
Moreover, another potential internal threat to the validity of our approach is the failure to take into account certain sensitive factors of the prompts, such as the number and order of examples. 
This could have an impact on the accuracy of the results obtained. 
Therefore, we plan to address this concern in future studies and explore the effect of these factors on the validity of our approach. 

\vspace{-1mm}
\subsection{External threat to the validity}
In terms of external threats, our current approach only employs API KG to guide LLM and evaluate the feasibility of KG-guided LLM.
However, to assess the generality of our approach, we plan to investigate other domains of KG-guided LLM. 
Unlike traditional clarification methods, our approach doesn't requires significant effort to maintain the database. 
However, extending our approach to other domains primarily entails replacing the KG of the corresponding domain and designing appropriate questioning aspects to guide the LLM effectively. 
Besides, the emergence of new LLMs, such as GPT-4~\cite{nori2023capabilities}, may have an impact on our approach. 
We are currently on the GPT-4 waiting list, we will utilize it to validate the effectiveness and generality of our approach in future.

\vspace{-1mm}
\subsection{Bridge the gap between KG and LLM}
The KG approach offers advantages such as efficient user interaction and accurate content generation aligned with human intentions. However, the KG approach relies on the knowledge graph, which can limit its fluency, and when unfamiliar topics arise, out-of-vocabulary (OOV) problems can occur.

In contrast, LLM-based approaches contain rich knowledge and can generate highly fluent chat content due to their in-context learning abilities. However, their broad knowledge base may introduce noise, making the LLM go off-topic when discussing familiar topics, and their lack of path planning ability can result in longer response times.

By combining KG and LLM, we can leverage the strengths of both approaches. We present a novel approach that uses a LLM guided by KG to achieve improved accuracy, efficiency, and fluency. Our approach opens a door to bridge the gap between KG and LLM, and we believe it can serve as a reference for future applications that use KG and LLM.

Currently, we have only demonstrated the effectiveness of our approach in the Java API domain. In the future, we will expand the scope of testing for our method and develop a practical tool.

\section{Related Work}
API recommendation~\cite{asaduzzaman2015exploring, he2021pyart, chen2021holistic, thung2016api} is an essential task in software engineering, and several methods~\cite{huang2018api, ling2019graph,liu2021opportunities, zhang2017expanding} have been proposed to improve its accuracy. Early techniques used query search methods~\cite{cao2021automated, groth2014api,ulrich2020analysis} that relied on literal and semantic matching of keywords between query statements and API descriptions, but they often failed to capture user intent due to insufficient keywords~\cite{zhang2018web,gu2016deep} in the query statements. Query extension methods~\cite{cao2021automated,lv2015codehow,gallant2019xu} were developed to supplement query information, but they lacked user involvement and failed to address the user intent problem.

Query clarification methods~\cite{cao2021automated,lv2015codehow,gallant2019xu,Eberhart2021DialogueMF} that used human-computer interaction were found to be more effective in capturing user needs. However, multiple rounds of question clarification~\cite{jannach2021survey} were often required, making the process inefficient. KG-based methods~\cite{huang2023answering, verborgh2016triple} were introduced to reduce the number of rounds by calculating the optimal questioning path for the API, but they were limited by the size of the knowledge base~\cite{huang2022prompt, le2021deep,karampatsis2020big} and the use of rigid templates to convert knowledge into questions and options~\cite{verborgh2016triple,morton2019robokop}.

To address these limitations, we propose a novel approach that leverages a LLM~\cite{brown2020language,openai2023gpt4,openai_GPT-3.5, wang2021codet5} guided by KG for API recommendation. Our approach overcomes OOV failures by using LLMs, which pre-train on the entire Internet corpus to provide a wealth of API knowledge. We filter out noise and transfer the optimal clarification path from the KG to the LLM to increase the accuracy of the clarification process~\cite{ren2021learning, basse2000clinical, datta2021generating}. Moreover, we use LLM as a neural knowledge base~\cite{huang2022prompt, petroni2019language, sharath2020question} to enhance the semantic understanding and in-context learning capabilities~\cite{wan2022they, huang2022prompt, huang2022se} of the system, generating fluent and appropriate clarification queries and options that address the limitations of the KG method in capturing user intent. Our approach has higher recommendation efficiency, accuracy, and fluency compared to previous KG-based methods.

\vspace{-2mm}
\section{Conclusion}
In this paper, we propose a novel knowledge-guided query clarification approach for API recommendation that leverages KG-guided LLM to address OOV failures and improve the fluency of conversations. 
By utilizing the LLM as neural knowledge base and taking advantage of its enhanced semantic understanding and in-context learning capabilities, our approach generates fluent and appropriate clarification questions and options. 
We also leverage the structured API knowledge and entity relationships in KG to filter out noise and transfer the optimal clarification path from KG to the LLM, increasing the efficiency of the clarification process.
Our approach is broken down into an AI chain that consists of five steps, each handled by a separate LLM call, resulting in improved accuracy, efficiency, and fluency for query clarification in API recommendation. 
Evaluation results demonstrate the effectiveness of our approach, outperforming the baselines regardless of the presence of extension keywords in the KG. 
Our approach bridges the gap between KG and LLM, combining the strengths and compensating for the weaknesses of both.
This presents a promising direction for future research in this area.
Our data can be found here\footnote{\href{https://anonymous.4open.science/r/AI-Chain-on-Knowledge-Guided-Pathfinding-for-Optimizing-Question-Clarification-in-API-Recommendation-179C/README.md}{{https://anonymous.4open.science/r/AI-Chain-on-Knowledge-Guided-Pathfinding-for-Optimizing-Question-Clarification-in-API-Recommendation-179C/README.md}}} .

\section{Acknowledgment}

The work is partly supported by the National Nature Science Foundation of China under Grant (Nos. 62262031), Cultivation Project for Academic and Technical Leader in Major Disciplines in Jiangxi Province (20232BCJ22013), and the Science and Technology Key Project of the Education Department of Jiangxi Province (GJJ2200302).

\bibliography{sample-base}

\begin{thebibliography}{10}

\bibitem{huang2018api}
Qiao Huang, Xin Xia, Zhenchang Xing, David Lo, and Xinyu Wang.
\newblock Api method recommendation without worrying about the task-api knowledge gap.
\newblock In {\em Proceedings of the 33rd ACM/IEEE International Conference on Automated Software Engineering}, pages 293--304, 2018.

\bibitem{stylos2006mica}
Jeffrey Stylos and Brad~A Myers.
\newblock Mica: A web-search tool for finding api components and examples.
\newblock In {\em Visual Languages and Human-Centric Computing (VL/HCC'06)}, pages 195--202. IEEE, 2006.

\bibitem{ling2019graph}
Chun-Yang Ling, Yan-Zhen Zou, Ze-Qi Lin, and Bing Xie.
\newblock Graph embedding based api graph search and recommendation.
\newblock {\em Journal of Computer Science and Technology}, 34:993--1006, 2019.

\bibitem{bajracharya2010searching}
Sushil Bajracharya, Joel Ossher, and Cristina Lopes.
\newblock Searching api usage examples in code repositories with sourcerer api search.
\newblock In {\em Proceedings of 2010 ICSE workshop on search-driven development: users, infrastructure, tools and evaluation}, pages 5--8, 2010.

\bibitem{Thung2013AutomaticRO}
Ferdian Thung, Shaowei Wang, D.~Lo, and Julia~L. Lawall.
\newblock Automatic recommendation of api methods from feature requests.
\newblock {\em 2013 28th IEEE/ACM International Conference on Automated Software Engineering (ASE)}, pages 290--300, 2013.

\bibitem{Rahman2016RACKAA}
Mohammad~Masudur Rahman, Chanchal~Kumar Roy, and D.~Lo.
\newblock Rack: Automatic api recommendation using crowdsourced knowledge.
\newblock {\em 2016 IEEE 23rd International Conference on Software Analysis, Evolution, and Reengineering (SANER)}, 1:349--359, 2016.

\bibitem{liu2021opportunities}
Chao Liu, Xin Xia, David Lo, Cuiyun Gao, Xiaohu Yang, and John Grundy.
\newblock Opportunities and challenges in code search tools.
\newblock {\em ACM Computing Surveys (CSUR)}, 54(9):1--40, 2021.

\bibitem{zhang2017expanding}
Feng Zhang, Haoran Niu, Iman Keivanloo, and Ying Zou.
\newblock Expanding queries for code search using semantically related api class-names.
\newblock {\em IEEE Transactions on Software Engineering}, 44(11):1070--1082, 2017.

\bibitem{cao2021automated}
Kaibo Cao, Chunyang Chen, Sebastian Baltes, Christoph Treude, and Xiang Chen.
\newblock Automated query reformulation for efficient search based on query logs from stack overflow.
\newblock In {\em 2021 IEEE/ACM 43rd International Conference on Software Engineering (ICSE)}, pages 1273--1285. IEEE, 2021.

\bibitem{eberhart2022generating}
Zachary Eberhart and Collin McMillan.
\newblock Generating clarifying questions for query refinement in source code search.
\newblock In {\em 2022 IEEE International Conference on Software Analysis, Evolution and Reengineering (SANER)}, pages 140--151. IEEE, 2022.

\bibitem{huang2023answering}
Qing Huang, Zishuai Li, Zhenchang Xing, Zhengkang Zuo, Xin Peng, Xiwei Xu, and Qinghua Lu.
\newblock Answering uncertain, under-specified api queries assisted by knowledge-aware human-ai dialogue.
\newblock {\em arXiv preprint arXiv:2304.14163}, 2023.

\bibitem{Huang2022112PK}
Qing Huang, Zhiqiang Yuan, Zhenchang Xing, Zhengkang Zuo, Changjing Wang, and Xin Xia.
\newblock 1+1>2: Programming know-what and know-how knowledge fusion, semantic enrichment and coherent application.
\newblock {\em ArXiv}, abs/2207.05560, 2022.

\bibitem{hssina2014comparative}
Badr Hssina, Abdelkarim Merbouha, Hanane Ezzikouri, and Mohammed Erritali.
\newblock A comparative study of decision tree id3 and c4. 5.
\newblock {\em International Journal of Advanced Computer Science and Applications}, 4(2):13--19, 2014.

\bibitem{brown2020language}
Tom Brown, Benjamin Mann, Nick Ryder, Melanie Subbiah, Jared~D Kaplan, Prafulla Dhariwal, Arvind Neelakantan, Pranav Shyam, Girish Sastry, Amanda Askell, et~al.
\newblock Language models are few-shot learners.
\newblock {\em Advances in neural information processing systems}, 33:1877--1901, 2020.

\bibitem{luccioni2021s}
Alexandra Luccioni and Joseph Viviano.
\newblock What’s in the box? an analysis of undesirable content in the common crawl corpus.
\newblock In {\em Proceedings of the 59th Annual Meeting of the Association for Computational Linguistics and the 11th International Joint Conference on Natural Language Processing (Volume 2: Short Papers)}, pages 182--189, 2021.

\bibitem{wan2022they}
Yao Wan, Wei Zhao, Hongyu Zhang, Yulei Sui, Guandong Xu, and Hai Jin.
\newblock What do they capture? a structural analysis of pre-trained language models for source code.
\newblock In {\em Proceedings of the 44th International Conference on Software Engineering}, pages 2377--2388, 2022.

\bibitem{huang2022prompt}
Qing Huang, Zhiqiang Yuan, Zhenchang Xing, Xiwei Xu, Liming Zhu, and Qinghua Lu.
\newblock Prompt-tuned code language model as a neural knowledge base for type inference in statically-typed partial code.
\newblock In {\em 37th IEEE/ACM International Conference on Automated Software Engineering}, pages 1--13, 2022.

\bibitem{huang2022se}
Qing Huang, Dianshu Liao, Zhenchang Xing, Zhiqiang Yuan, Qinghua Lu, Xiwei Xu, and Jiaxing Lu.
\newblock Se factual knowledge in frozen giant code model: A study on fqn and its retrieval.
\newblock {\em arXiv preprint arXiv:2212.08221}, 2022.

\bibitem{nori2023capabilities}
Harsha Nori, Nicholas King, Scott~Mayer McKinney, Dean Carignan, and Eric Horvitz.
\newblock Capabilities of gpt-4 on medical challenge problems.
\newblock {\em arXiv preprint arXiv:2303.13375}, 2023.

\bibitem{lee2022coauthor}
Mina Lee, Percy Liang, and Qian Yang.
\newblock Coauthor: Designing a human-ai collaborative writing dataset for exploring language model capabilities.
\newblock In {\em Proceedings of the 2022 CHI Conference on Human Factors in Computing Systems}, pages 1--19, 2022.

\bibitem{huang20221+}
Qing Huang, Zhiqiang Yuan, Zhenchang Xing, Zhengkang Zuo, Changjing Wang, and Xin Xia.
\newblock 1+ 1$> $2: Programming know-what and know-how knowledge fusion, semantic enrichment and coherent application.
\newblock {\em IEEE Transactions on Services Computing}, 2022.

\bibitem{li2018improving}
Hongwei Li, Sirui Li, Jiamou Sun, Zhenchang Xing, Xin Peng, Mingwei Liu, and Xuejiao Zhao.
\newblock Improving api caveats accessibility by mining api caveats knowledge graph.
\newblock In {\em 2018 IEEE International Conference on Software Maintenance and Evolution (ICSME)}, pages 183--193. IEEE, 2018.

\bibitem{liu2020generating}
Yang Liu, Mingwei Liu, Xin Peng, Christoph Treude, Zhenchang Xing, and Xiaoxin Zhang.
\newblock Generating concept based api element comparison using a knowledge graph.
\newblock In {\em Proceedings of the 35th IEEE/ACM International Conference on Automated Software Engineering}, pages 834--845, 2020.

\bibitem{blackwell2000cognitive}
Alan~F Blackwell and Thomas~RG Green.
\newblock A cognitive dimensions questionnaire optimised for users.
\newblock In {\em PPIG}, volume~13. Citeseer, 2000.

\bibitem{Radev2002EvaluatingWQ}
Dragomir~R. Radev, Hong Qi, Harris Wu, and Weiguo Fan.
\newblock Evaluating web-based question answering systems.
\newblock In {\em International Conference on Language Resources and Evaluation}, 2002.

\bibitem{Sanderson2010ChristopherDM}
Mark Sanderson.
\newblock Christopher d. manning, prabhakar raghavan, hinrich sch{\"u}tze, introduction to information retrieval, cambridge university press 2008. isbn-13 978-0-521-86571-5, xxi + 482 pages.
\newblock {\em Natural Language Engineering}, 16:100 -- 103, 2010.

\bibitem{asaduzzaman2015exploring}
Muhammad Asaduzzaman, Chanchal~K Roy, Samiul Monir, and Kevin~A Schneider.
\newblock Exploring api method parameter recommendations.
\newblock In {\em 2015 IEEE International Conference on Software Maintenance and Evolution (ICSME)}, pages 271--280. IEEE, 2015.

\bibitem{he2021pyart}
Xincheng He, Lei Xu, Xiangyu Zhang, Rui Hao, Yang Feng, and Baowen Xu.
\newblock Pyart: Python api recommendation in real-time.
\newblock In {\em 2021 IEEE/ACM 43rd International Conference on Software Engineering (ICSE)}, pages 1634--1645. IEEE, 2021.

\bibitem{chen2021holistic}
Chi Chen, Xin Peng, Zhenchang Xing, Jun Sun, Xin Wang, Yifan Zhao, and Wenyun Zhao.
\newblock Holistic combination of structural and textual code information for context based api recommendation.
\newblock {\em IEEE Transactions on Software Engineering}, 48(8):2987--3009, 2021.

\bibitem{thung2016api}
Ferdian Thung.
\newblock Api recommendation system for software development.
\newblock In {\em Proceedings of the 31st IEEE/ACM International Conference on Automated Software Engineering}, pages 896--899, 2016.

\bibitem{groth2014api}
Paul Groth, Antonis Loizou, Alasdair~JG Gray, Carole Goble, Lee Harland, and Steve Pettifer.
\newblock Api-centric linked data integration: The open phacts discovery platform case study.
\newblock {\em Journal of web semantics}, 29:12--18, 2014.

\bibitem{ulrich2020analysis}
Hannes Ulrich, Ann-Kristin Kock-Schoppenhauer, Cora Drenkhahn, Matthias L{\"o}be, and Josef Ingenerf.
\newblock Analysis of iso/ts 21526 towards the extension of a standardized query api.
\newblock In {\em Integrated Citizen Centered Digital Health and Social Care}, pages 202--206. IOS Press, 2020.

\bibitem{zhang2018web}
Neng Zhang, Jian Wang, Yutao Ma, Keqing He, Zheng Li, and Xiaoqing~Frank Liu.
\newblock Web service discovery based on goal-oriented query expansion.
\newblock {\em Journal of Systems and Software}, 142:73--91, 2018.

\bibitem{gu2016deep}
Xiaodong Gu, Hongyu Zhang, Dongmei Zhang, and Sunghun Kim.
\newblock Deep api learning.
\newblock In {\em Proceedings of the 2016 24th ACM SIGSOFT international symposium on foundations of software engineering}, pages 631--642, 2016.

\bibitem{lv2015codehow}
Fei Lv, Hongyu Zhang, Jian-guang Lou, Shaowei Wang, Dongmei Zhang, and Jianjun Zhao.
\newblock Codehow: Effective code search based on api understanding and extended boolean model (e).
\newblock In {\em 2015 30th IEEE/ACM International Conference on Automated Software Engineering (ASE)}, pages 260--270. IEEE, 2015.

\bibitem{gallant2019xu}
Morgan Gallant, Haruna Isah, Farhana Zulkernine, and Shahzad Khan.
\newblock Xu: an automated query expansion and optimization tool.
\newblock In {\em 2019 IEEE 43rd Annual Computer Software and Applications Conference (COMPSAC)}, volume~1, pages 443--452. IEEE, 2019.

\bibitem{Eberhart2021DialogueMF}
Zachary Eberhart and Collin McMillan.
\newblock Dialogue management for interactive api search.
\newblock {\em 2021 IEEE International Conference on Software Maintenance and Evolution (ICSME)}, pages 274--285, 2021.

\bibitem{jannach2021survey}
Dietmar Jannach, Ahtsham Manzoor, Wanling Cai, and Li~Chen.
\newblock A survey on conversational recommender systems.
\newblock {\em ACM Computing Surveys (CSUR)}, 54(5):1--36, 2021.

\bibitem{verborgh2016triple}
Ruben Verborgh, Miel Vander~Sande, Olaf Hartig, Joachim Van~Herwegen, Laurens De~Vocht, Ben De~Meester, Gerald Haesendonck, and Pieter Colpaert.
\newblock Triple pattern fragments: a low-cost knowledge graph interface for the web.
\newblock {\em Journal of Web Semantics}, 37:184--206, 2016.

\bibitem{le2021deep}
Duc Le, Gil Keren, Julian Chan, Jay Mahadeokar, Christian Fuegen, and Michael~L Seltzer.
\newblock Deep shallow fusion for rnn-t personalization.
\newblock In {\em 2021 IEEE Spoken Language Technology Workshop (SLT)}, pages 251--257. IEEE, 2021.

\bibitem{karampatsis2020big}
Rafael-Michael Karampatsis, Hlib Babii, Romain Robbes, Charles Sutton, and Andrea Janes.
\newblock Big code!= big vocabulary: Open-vocabulary models for source code.
\newblock In {\em Proceedings of the ACM/IEEE 42nd International Conference on Software Engineering}, pages 1073--1085, 2020.

\bibitem{morton2019robokop}
Kenneth Morton, Patrick Wang, Chris Bizon, Steven Cox, James Balhoff, Yaphet Kebede, Karamarie Fecho, and Alexander Tropsha.
\newblock Robokop: an abstraction layer and user interface for knowledge graphs to support question answering.
\newblock {\em Bioinformatics}, 35(24):5382--5384, 2019.

\bibitem{openai2023gpt4}
OpenAI.
\newblock Gpt-4 technical report, 2023.

\bibitem{openai_GPT-3.5}
OpenAI.
\newblock Openai gpt-3.5.
\newblock \url{https://openai.com/blog/introducing-chatgpt-and-whisper-apis}, 2023.
\newblock Accessed: 2023.

\bibitem{wang2021codet5}
Yue Wang, Weishi Wang, Shafiq Joty, and Steven~CH Hoi.
\newblock Codet5: Identifier-aware unified pre-trained encoder-decoder models for code understanding and generation.
\newblock {\em arXiv preprint arXiv:2109.00859}, 2021.

\bibitem{ren2021learning}
Xuhui Ren, Hongzhi Yin, Tong Chen, Hao Wang, Zi~Huang, and Kai Zheng.
\newblock Learning to ask appropriate questions in conversational recommendation.
\newblock In {\em Proceedings of the 44th international ACM SIGIR conference on research and development in information retrieval}, pages 808--817, 2021.

\bibitem{basse2000clinical}
Linda Basse, Dorthe~Hjort Jakobsen, Per Billesb{\o}lle, Mads Werner, and Henrik Kehlet.
\newblock A clinical pathway to accelerate recovery after colonic resection.
\newblock {\em Annals of surgery}, 232(1):51, 2000.

\bibitem{datta2021generating}
Soham Datta, Prabir Mallick, Sangameshwar Patil, Indrajit Bhattacharya, and Girish Palshikar.
\newblock Generating an optimal interview question plan using a knowledge graph and integer linear programming.
\newblock In {\em Proceedings of the 2021 Conference of the North American Chapter of the Association for Computational Linguistics: Human Language Technologies}, pages 1996--2005, 2021.

\bibitem{petroni2019language}
Fabio Petroni, Tim Rockt{\"a}schel, Patrick Lewis, Anton Bakhtin, Yuxiang Wu, Alexander~H Miller, and Sebastian Riedel.
\newblock Language models as knowledge bases?
\newblock {\em arXiv preprint arXiv:1909.01066}, 2019.

\bibitem{sharath2020question}
Japa~Sai Sharath and Rekabdar Banafsheh.
\newblock Question answering over knowledge base using language model embeddings.
\newblock In {\em 2020 International Joint Conference on Neural Networks (IJCNN)}, pages 1--8. IEEE, 2020.

\end{thebibliography}
\end{document}